# Effective Mass in Bilayer Graphene at Low Carrier Densities: the Role of Potential Disorder and Electron-Electron Interaction

J. Li[1], L. Z. Tan[2,3]†*, K. Zou[1]‡, A. A. Stabile[1], D. J. Seiwell[1], K. Watanabe[4], T. Taniguchi[4], Steven G. Louie[2,3]*, J. Zhu[1,5]*

[1]*Department of Physics, The Pennsylvania State University, University Park, Pennsylvania 16802, USA*

[2]*Department of Physics, University of California at Berkeley, Berkeley, California 94720, USA*

[3]*Materials Sciences Division, Lawrence Berkeley National Laboratory, Berkeley, California 94720, USA*

[4]*National Institute for Material Science, 1-1 Namiki, Tsukuba 305-0044, Japan*

[5]*Center for 2-Dimensional and Layered Materials, The Pennsylvania State University, University Park, Pennsylvania 16802, USA*

†*Current address: The Makineni Theoretical Laboratories, Department of Chemistry, University of Pennsylvania, Philadelphia, Pennsylvania 19104-6323, USA*

‡*Current address: Department of Applied Physics and Center for Research on Interface Structures and Phenomena (CRISP), Yale University, New Haven, Connecticut 06520, USA*

*Corresponding to: jzhu@phys.psu.edu (J. Zhu), sglouie@berkeley.edu (S. G. Louie), and liangtan@sas.upenn.edu (L. Z. Tan)

**Abstract**

In a two-dimensional electron gas, the electron-electron interaction generally becomes stronger at lower carrier densities and renormalizes the Fermi liquid parameters such as the effective mass of carriers. We combine experiment and theory to study the effective masses of electrons and holes $m^*_e$ and $m^*_h$ in bilayer graphene in the low carrier density regime of order $1 \times 10^{11}$ cm$^{-2}$. Measurements use temperature-dependent low-field Shubnikov-de Haas (SdH) oscillations are observed in high-mobility hexagonal boron nitride (h-BN) supported samples. We find that while $m^*_e$ follows a tight-binding description in the whole density range, $m^*_h$ starts to drop rapidly below the tight-binding description at carrier density $n = 6 \times 10^{11}$ cm$^{-2}$ and exhibits a strong suppression of 30% when $n$ reaches $2 \times 10^{11}$ cm$^{-2}$. Contributions from electron-electron interaction alone, evaluated using several different approximations, cannot explain the experimental trend. Instead, the effect of potential fluctuation and the resulting electron-hole puddles play a crucial role. Calculations including both the electron-electron interaction and disorder effects explain the experimental data qualitatively and quantitatively. This study reveals an unusual disorder effect unique to two-dimensional semi-metallic systems.

Bilayer graphene is a unique two-dimensional electron gas (2DEG) system with unusual electronic properties [1]. At high carrier densities, its hyperbolic bands are well described by a four-band Hamiltonian [2, 3] given by the tight-binding (TB) description [4], where the hopping parameters are determined by experiments or first-principles calculations [5-10]. Close to the charge neutrality point (CNP), bilayer graphene exhibits fascinating electron-electron (e-e) interaction driven ground states [11-15]. A natural question arises: How does the density of states of bilayer graphene at the Fermi energy evolve as carrier density $n$ decreases continuously? The study of the effective carrier mass $m^*$ is a powerful tool to probe this evolution. Indeed, in conventional 2DEGs, increasing e-e interaction leads to substantial increase of $m^*$ at low carrier densities, long before predicated many-body instabilities [16-21]. Such studies provide valuable inputs to advance many-body calculations [22]. In monolayer and bilayer graphene, the close proximity of the conduction and valence bands and their pseudospin characters, play a significant role in the screening of the Coulomb interaction. This has consequences for the dispersions of the elementary excitations and the transport properties of these systems [23-26]. In monolayer graphene, both calculations[27], and measurements of $m^*$[28] [29] report strong enhancement of the Fermi velocity $v_F$ at low carrier densities. The situation in bilayer graphene is much less clear. Existing theoretical predictions vary greatly on the sign and magnitude of the interaction correction to $m^*$ [30-35] while measurements have been lacking.

In our earlier work [10], we reported on the measurements of $m^*$ in bilayer graphene in the density regime of order $1 \times 10^{12}$ cm$^{-2}$. A TB description was found to work well, the hopping parameters of which were accurately extracted from data. As the previous samples rested on oxides, Coulomb potential disorder (field effect mobility $\mu_{FE}$ ~ a few thousand cm$^2$V$^{-1}$s$^{-1}$ and disorder energy $\delta E$ of a few tens of meV [36, 37]) prevented measurements at lower densities. In our current h-BN supported samples, $\mu_{FE}$ reaches 30,000 cm$^2$V$^{-1}$s$^{-1}$, which allows for precise determination of $m^*$ down to $n = 2 \times 10^{11}$ cm$^{-2}$ for both electrons and holes. Following the conventional definition of the interaction parameter $r_s = U/E_F$, where $U$ is the Coulomb interaction energy $e^2\sqrt{n\pi}/(4\pi\varepsilon_0\varepsilon)$ and $E_F$ is the Fermi energy, we estimate $r_s$ to be $7.5/\sqrt{n(\text{in unit of } 10^{11}\text{cm}^{-2})}$ using $m^* = 0.033$ m$_e$, which is the average value of the measured electron and hole masses near $1 \times 10^{12}$ cm$^{-2}$ in Ref. [10]. In our presently studied carrier density regime ($2 - 12 \times 10^{11}$ cm$^{-2}$), $r_s$ ranges from 2.2 to 5.3, which is comparable to the range studied in GaAs electron 2DEG, where the renormalized $m^*$ exceeds the band mass by 40% at $r_s$ ~ 5 due to e-e interaction [18]. Here, we find that $m^*_e$ and $m^*_h$ behave very differently as $n$ decreases. While $m^*_e$ continues to follow the high-density TB extrapolation, $m^*_h$ sharply dives in value below $n = 6 \times 10^{11}$ cm$^{-2}$, reaching about 70% of the TB band mass at $n = 2 \times 10^{11}$ cm$^{-2}$. A thorough theoretical investigation evaluating the effect of e-e interaction in different approximations, together with the effect of Coulomb potential disorder, identifies density inhomogeneity to be a key factor in explaining the experimental observations. This unusual effect of disorder is unique to 2D semi-metallic systems.

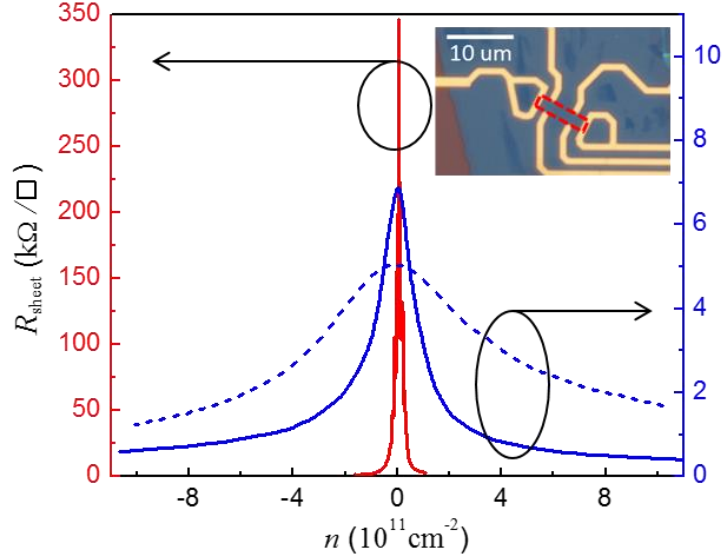

FIG. 1. Sheet resistance vs carrier density $R_{sheet}(n)$ for samples A (solid red), B (solid blue) and C (dashed blue). Samples A and B are supported on h-BN, sample C on SiO$_2$. The field effect mobility $\mu_{FE}$ is 30,000 cm$^2$V$^{-1}$s$^{-1}$, 22,000 cm$^2$V$^{-1}$s$^{-1}$, and 4000 cm$^2$V$^{-1}$s$^{-1}$ respectively for samples A to C. $T = 1.6$ K. The large resistance sample A exhibits at the CNP results from a finite band gap caused by unintentional doping. We discuss the effect of a band gap on the band mass in S4 of the supplementary material [39]. Inset: An optical micrograph for sample A.

Bilayer multi-terminal devices are made by exfoliating, transferring, stacking and patterning of multi-layer-graphene bottom gate electrode, 15 – 30nm thick h-BN gate dielectric (Momentive, Polartherm grade PT110 and NIMS) and bilayer graphene sheet (Kish Graphite) using a PMMA/PVA based transfer method [38] and standard e-beam lithography. Transport experiments are carried out in a variable-temperature, pumped He$^4$ cryostat with a 9 T magnet using standard low-frequency lock-in technique (47 Hz) with current excitation 50 nA. Figure 1 plots the sheet resistance vs carrier density $R_{sheet}(n)$ of samples A and B, together with sample C reported in Zou et al [10] for comparison. The field effect mobility $\mu_{FE}$ is 30,000 cm$^2$V$^{-1}$s$^{-1}$ and 22,000 cm$^2$V$^{-1}$s$^{-1}$ respectively in samples A and B, in comparison to $\mu_{FE}$ = 4,000 cm$^2$V$^{-1}$s$^{-1}$ in sample C, which is supported on SiO$_2$ substrate. The unintentional doping for both devices are moderate, and the effect of the displacement ($D$) field on the bare band mass is modeled in S4 of the supplementary material for both devices [39]. We find that the presence of a small $D$-field does not change the conclusions of the paper.

The effective mass $m^*$ as measured in quantum oscillations is given by

$$m^* = \frac{\hbar^2}{2\pi} \frac{dA(E)}{dE}\bigg|_{E=E_F} \quad (1)$$

where $A(E)$ is the $k$-space area enclosed by the contour of constant energy $E$ in the quasi-particle band structure. To accurately determine $m^*$, we measure the temperature-dependent magneto-resistance $R_{xx}(B)$ at a fixed carrier density (Fig. 2(a)), extract the low-field Shubnikov de Haas (SdH) oscillation amplitude $\delta R_{xx}(T, B)$ and perform simultaneous fitting of the temperature and magnetic field dependence to the Lifshitz-Kosevich formula[40],

$$\frac{\delta R_{xx}}{R_0} = 4\gamma_{th}\exp\left(\frac{-\pi}{\omega_c\tau_q}\right), \gamma_{th}=\frac{2\pi^2 k_B T/\hbar\omega_c}{\sinh(2\pi^2 k_B T/\hbar\omega_c)} \quad (2)$$

where $\omega_c = \frac{eB}{m^*}$ is the cyclotron frequency. The effective mass $m^*$ and the quantum scattering time $\tau_q$ are the two fitting parameters.

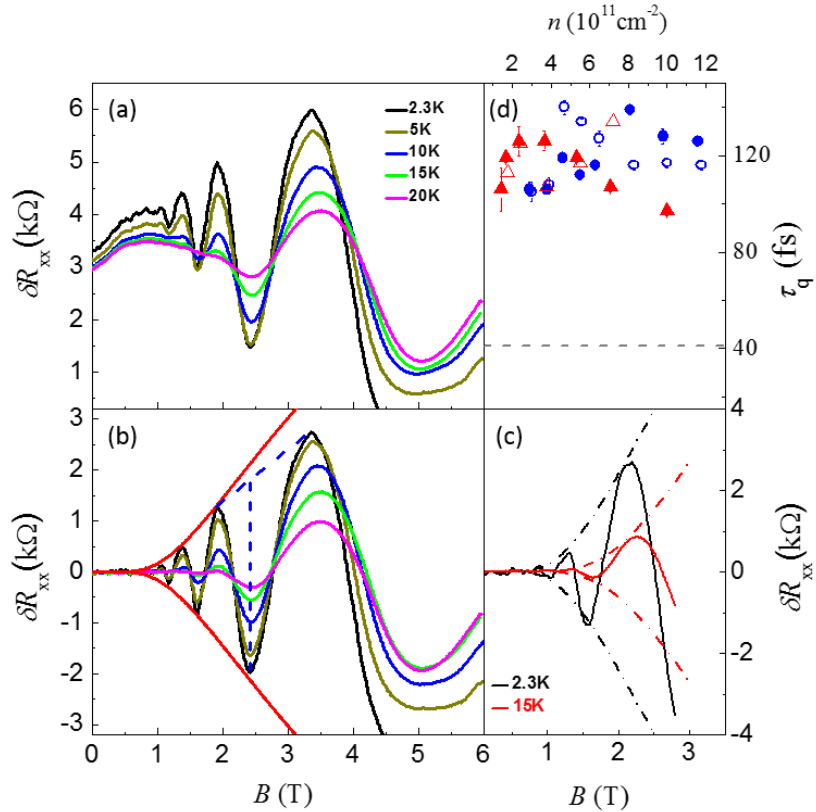

FIG. 2. (a) $T$-dependent magnetoresistance $R_{xx}(B)$ for $n_h = 4.7 \times 10^{11}$ cm$^{-2}$ at selected temperatures as indicated in the plot. (b) Oscillation amplitude $\delta R_{xx}(B)$ of data in (a) after background subtraction. The solid red curve plots Eq.(2) with fitting parameters $m_h^* = 0.0347$ $m_e$ and $\tau_q = 140$ fs. $T = 2.3$ K. $\delta R_{xx}(B)$ starts deviating from the fit above $B = 3$ T. Conventional method used to extract $\delta R_{xx}$ is illustrated by the blue dashed lines and produces $m^* = 0.0311(2)$ $m_e$. This is 10% smaller than $m_h^* = 0.0347$ $m_e$ obtained from the global fitting. (c) $\delta R_{xx}(B)$ for $n_h = 3.0 \times 10^{11}$ cm$^{-2}$ at $T = 2.3$ K and $T = 15$ K. Dashed curves are fits to Eq.(2) with $m_h^* = 0.0285$ $m_e$ and $\tau_q = 107$ fs. Data in (a)-(c) are from sample B. (d) The quantum scattering time $\tau_q$ as a function of carrier density in sample A (red symbols) and sample B (blue symbols). Electrons are shown in filled symbols and holes in open symbols. $\tau_q$ is about 40 fs (dashed grey line) in sample C (Ref. [10]).

This global fitting procedure is illustrated in Figs. 2(b) and (c) for two carrier densities $n_h$ = 4.7 and 3.0 × 10$^{11}$ cm$^{-2}$ as examples (see S1 and S2 of the supplementary material [39]). Compared to common practice of approximating $\delta R_{xx}$ at a particular $B$-field by linearly interpolating adjacent peak heights and analyze its $T$-dependence to obtain $m^*$, fits to Eq. (2) better represent the oscillation amplitude $\delta R_{xx}$, especially at low carrier densities when only a few oscillations are available (See Fig. 2(c) for example). It also enables us to discern and avoid using the $T$-dependent oscillations of nascent quantum Hall states, the analysis of which can lead to error in $m^*$ (see caption in Fig. 2(b)). The effective mass $m^*$ obtained using the global fitting procedure is independent of the $B$ field by virtue of the method and best extrapolates to the density-of-states mass of the bilayer graphene at $B = 0$, which is expected to be modified by e-e interactions [30-35].

The above analysis enables us to accurately determine both the electron and hole effective mass $m^*_h$ and $m^*_e$ for the approximate carrier density range of 1 - 10 × 10$^{11}$cm$^{-2}$. The uncertainty of $m^*$ varies from ± 0.0002 $m_e$ to ± 0.004 $m_e$ from high to low densities. The high accuracy of the measurements facilitates comparison to theory as interaction corrections to $m^*$ are expected to be typically in the few to tens of percent range [16, 18]. Also plotted in Fig. 2(d) is the quantum scattering time $\tau_q$ in both samples. $\tau_q$ is between 100 and 140 fs for both electrons and holes. Compared to ~ 40 fs in sample C [10], the high values of $\tau_q$ in samples A and B attest to the improvement of sample quality. Below $n$ =1 × 10$^{11}$ cm$^{-2}$, the SdH oscillations become increasingly more non-sinusoidal due to density inhomogeneity and global fits cannot be obtained reliably.

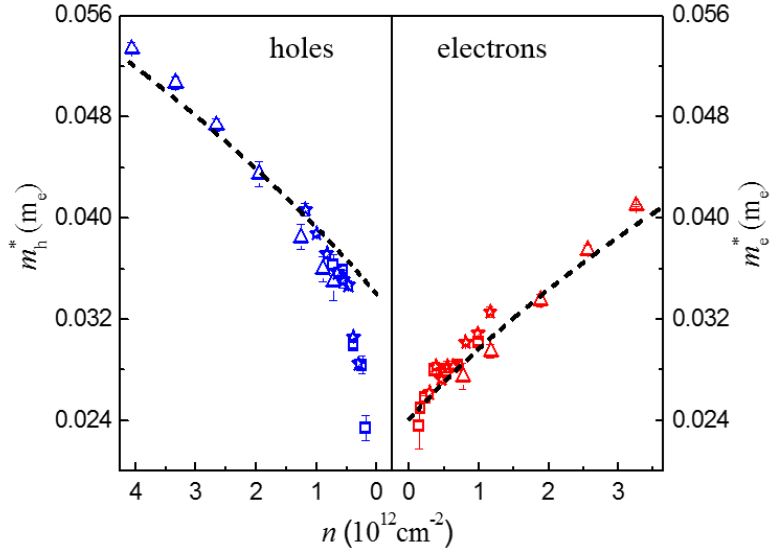

FIG. 3. The effective carrier mass $m_h^*$ and $m_e^*$ as a function of the carrier density (red for electrons, blue for holes) in samples A (squares), B (stars), and C (triangles). Data on C is from Ref. [10]. Together, the measurement covers the density range of approximately 1.4 - 41 × 10$^{11}$ cm$^{-2}$. The dashed curves plot $m^*$ calculated using a 4 × 4 tight-binding Hamiltonian with hopping parameters $\gamma_0$ = 3.43 eV, $\gamma_1$ = 0.40 eV, $\gamma_3$ = 0, and $v_4$ = 0.063. These values are obtained in Ref.[10] by fitting the data in sample C at high densities.

Figure 3 plots $m^*_h$ and $m^*_e$ obtained in samples A and B, together with data from sample C in Ref. [10]. In the overlapping density regime, current and previous results agree very well and are well described by the TB model with hopping parameters $\gamma_0 = 3.43$ eV, $\gamma_1 = 0.40$ eV $\gamma_3 = 0$ and $v_4 = \gamma_4/\gamma_0 = 0.063$, $\Delta = 0.018$ eV, which are determined in Ref. [10]. The calculated $m^*$ are plotted as dashed lines in Fig. 3. The electron and hole branches use the same set of parameters, with their mass differences captured by $v_4$. In current samples, the TB parameters continue to describe all the $m^*_e$ data very well down to the lowest density measured. On the hole side, however, $m^*_h$ exhibits a sharp drop from the TB model as $n_h$ is decreased to less than $5 \times 10^{11}$ cm$^{-2}$, reaching a large suppression of 30% at $n_h = 2 \times 10^{11}$ cm$^{-2}$. These densities are still sufficiently high that the effect of trigonal warping [1] can be safely neglected. (Fig. S6 of the supplementary material [39])

In existing theoretical studies of bilayer electronic dispersions, the effect of e-e interaction manifests in two ways, *i. e.* by renormalizing the hopping parameters within the TB model at high carrier densities [33] and by causing deviations of $m^*$ from the TB description at low carrier densities. There different trends of $m^*$ predicted [30-32, 34, 35].

We begin our calculations with a four-band TB Hamiltonian with non-interacting hopping parameters and explicitly include e-e interaction with the random phase approximation (RPA) of the screened exchange self-energy

$$\Sigma(k) = -\sum_q \frac{V^{2D}(q)}{\varepsilon(q)} F^{ss'}(k, k+q) \qquad (3)$$

using a dielectric function $\varepsilon(q) = \varepsilon_{BN} - V^{2D}(q)\chi(q)$, that includes contributions from both the bilayer graphene and the h-BN substrate and overlayer. Here $\varepsilon_{BN} = 3.0$ is determined from the gating efficiency of the backgate, and $F^{ss'}$ is the pseudospin overlap factor [30, 31]. Eq. (3) provides the RPA correction to the bare energy bands $E_0(k)$ obtained from TB calculation to yield the quasiparticle band structure $E(k) = E_0(k) + \Sigma(k)$. The effective mass is then computed using Eq. (1).

The calculated $m^*_e$ and $m^*_h$ are plotted in Fig. 4 in olive dotted lines. Interaction leads to a slightly faster decrease of $m^*_e$ and $m^*_h$ at low carrier densities, in contrast to the sudden drop observed in the measured $m^*_h$ for $n_h < 5 \times 10^{11}$ cm$^{-2}$. Examining the problem from a different angle, we note that in the RPA model, the dielectric function is well described by the Thomas-Fermi (TF) screening $\varepsilon(q) = \varepsilon_{BN} + \frac{q_{TF}}{q}$ in the small $q$ limit [34]. Fitting the TF description to our data yields a ten-fold reduction of the TF screening wavevector $q_{TF}$ from its expected value of $q_{TF} = m^*e^2/\hbar^2$. This would imply extremely weak screening of the e-e interaction in our devices, which cannot be justified. (see Fig. S7 of the supplementary material [39]). Thus, e-e interaction effect, at least at the RPA level, appears to be too weak to account for the experimental observations. In comparison, in monolayer graphene, a large suppression of $m^*$ is also observed at low carrier densities and well described by RPA calculations [28].

Can Coulomb potential fluctuation and the resulting density inhomogeneity[36, 37, 41] play a role? The answer is not so intuitive at the first glance. In a conventional semiconducting 2DEG, density inhomogeneity results in the smearing of $m^*(n)$. This effect does not alter the trend of $m^*(n)$ and is typically non-consequential in the carrier density regime where the SdH oscillations are well-behaved. In Fig. 2(c)), the SdH oscillations at $n_h = 3 \times 10^{11}$ cm$^{-2}$ appear to be well-behaved, yet the measured $m^*_h$ is already 14% below the TB band mass. Here, the *gapless* nature of the bilayer bands makes a crucial difference between bilayer graphene and a conventional 2DEG. As the inset of Fig. 4 illustrates, as the Fermi energy $E_F$ approaches the disorder energy scale $\delta E$, instead of depletion, carriers of the opposite sign start to appear in parts of the sample. The SdH oscillations of a minority carrier type have the opposite sign in $dA/dE$; their presence in some regions of the sample thus contribute negatively to the average of $m^*$, resulting in a decrease in its value. Such cancellation effect does not occur in a conventional semiconductor 2DEG.

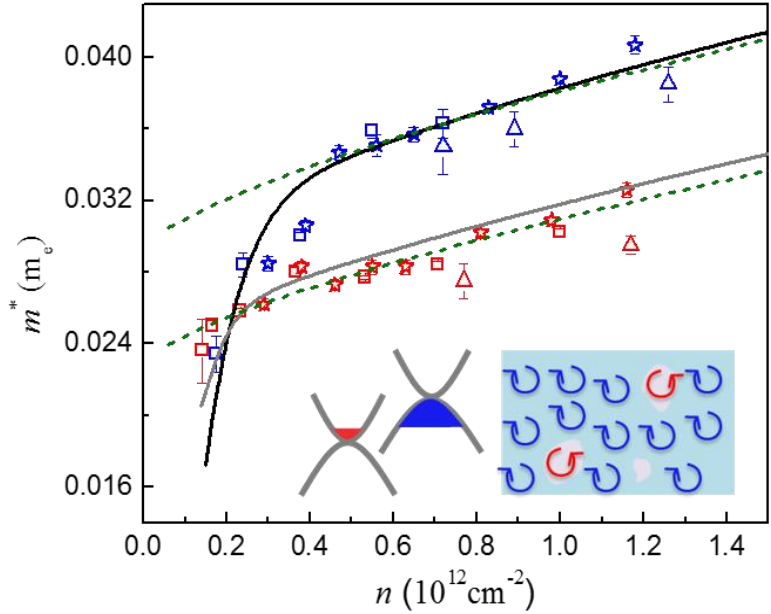

FIG. 4. Comparison of calculations and experiment at low carrier density (0.2 – 1.3 × 10$^{12}$ cm$^{-2}$). Experimental data follow the symbols used in Fig. 3. The olive dashed lines plot the calculated $m^*$ including e-e interaction in a random phase approximation. The black and gray lines are calculations that further include the effect of potential disorder using $\delta E = 5.4$ meV obtained from $\tau_q$ and the temperature dependence of the conductance. In both calculations, $\gamma_0 = 3.08$ eV and $\gamma_1 = 0.36$ eV are chosen to fit the experimental data in the high-density regime. Their values differ from those obtained in Ref. [10] since e-e interaction is explicitly calculated here whereas in Ref.[10] its effect is represented by renormalizing the hopping parameters. $\gamma_3 = 0$ and $v_4 = 0.063$ are taken from Ref. [10]. Inset: A schematic illustration of the electron-hole coexistence at low carrier densities due to disorder and its effect on the cyclotron motion.

This situation can be modeling by defining the overall carrier density and effective mass as ensemble averages of their local counterparts $n_{loc}$ and $m_{loc}$ respectively:

$$n(E) = \langle n_{loc} \rangle = \int d\mu\, f(\mu)\, n_{loc}(E + \mu) \quad (4)$$
$$m(E) = \langle m_{loc} \rangle = \int d\mu\, f(\mu)\, m_{loc}(E + \mu) \quad (5)$$

Here, the fluctuation of energy is assumed to have a Gaussian profile $f(\mu)$ with standard deviation $\delta E$.

Effective masses calculated using the RPA model and including disorder characterized by a broadening energy $\delta E = 5.4$ meV are plotted as solid lines in Fig. 4. Evidently, the combination of the e-e interaction and Coulomb potential fluctuations can now quantitatively reproduce the observed behavior of $m^*_e$ and $m^*_h$ over the entire range of measurement and for both samples. Remarkably, the same value for $\delta E$ simultaneously captures the sharp decrease of $m^*_h$ at $n_h < 5 \times 10^{11}$ cm$^{-2}$ and the absence of such decrease on the electron side. Our calculations predict that $m^*_e$ should also substantially decrease from the TB values at yet lower carrier densities, just below the range probed in our measurements. The difference arises from a smaller electron density inhomogeneity due to a smaller $m^*_e$. The quantum scattering time $\tau_q \sim 120$ fs found in both samples (Fig. 2(d)) yields $\delta E \sim \hbar/2\tau_q \sim 2.7$ meV, in good agreement with the theoretical fit. In addition, we can estimate the density fluctuation $\delta n$ by locating the onset density $n^*$ at which the conductance sharply increases with density [11-15]. $n^*$ is approximately $2 \times 10^{10}$ cm$^{-2}$ in sample A and $4 \times 10^{10}$ cm$^{-2}$ in sample B (Fig. S4). These values are also consistent with estimates obtained by locating the crossover density $n(h/e)_c \sim 5 \times 10^{10}$ cm$^{-2}$, where the temperature dependence of $R(n)$ changes from that of a metal, *i. e.* $dR/dT > 0$ to that of an insulator, *i. e.* $dR/dT < 0$ [42] in a bilayer sample of similar quality. A $\delta n$ of $5 \times 10^{10}$ cm$^{-2}$ corresponds to $\delta E = 2$ meV using $m^* = 0.03\, m_e$. These consistent estimates of disorder energy scales support the fitting value of $\delta E$ used for both samples. Furthermore, our calculations also show that interaction renormalizes the inter-band transition energy $\gamma_1$ from the "bare" value of 0.36 eV (Fig. 4) to 0.38 eV, in excellent agreement with infra-red absorption measurements [6, 7, 9].

In Ref. [10], we have shown that a set of renormalized TB hopping parameters can capture $m^*$ in the high-density regime very well, without explicitly including e-e interactions (See dashed lines in Fig. 3). In Fig. S8 of the supplementary material [39], we show that adding disorder broadening $\delta E$ to this set of parameters can also capture the main trend of data, with the diving of $m^*_h$ at low densities slightly too abrupt compared to experiment.

The above studies highlight a few remarkable differences between bilayer graphene, a gapless Dirac Fermi liquid and conventional semiconductor 2DEGs. Firstly, both our calculations and measurements suggest that the effect of e-e interaction on $m^*$ in bilayer graphene remains weak down to $n \sim 2 \times 10^{11}$ cm$^{-2}$ ($r_s = 5.3$) while past studies on GaAs electrons showed an enhancement of more than 40% at this interaction parameter [18]. Secondly, the effect of disorder appears quite different in these two systems. In conventional semiconducting 2DEGs, disorder leads to localization and therefore the *increase*, rather than the decrease of $m^*$ at low carrier densities [18]. Here in gapless bilayer graphene, disorder leads to coexisting electrons and holes and consequently a partial cancellation effect on $m^*$. In comparison to the well-recognized Klein tunneling effect in p-n junctions [43, 44], this study exposed a more elusive effect of electron-hole puddles. Studies of low-carrier-density regimes in Dirac materials thus require a great deal of caution. For now, samples of yet higher qualities are necessary to elucidate the intrinsic behavior of $m^*$ near the charge neutral point of bilayer graphene.

In conclusion, we have performed careful measurements of the effective mass $m^*$ in high-quality h-BN supported bilayer graphene samples down to the carrier density regime of $1 \times 10^{11}$ cm$^{-2}$ and observed sharp decrease of the hole mass at low carrier densities. Our calculations show that while the inclusion of electron-electron interaction is necessary to reach excellent quantitative agreement with data at all carrier densities, Coulomb potential fluctuations, which result in the co-existence of electron and hole regions and a partial cancellation of $m^*$, is chiefly responsible for the observed sharp drop in $m^*_h$ at low densities. This mechanism, which is absent in finite-gap semiconductor two-dimensional systems, is another manifestation of the unusual consequences of gapless Dirac bands.

**Acknowledgement**

J. L., K. Z., A. A. S. D. J. S. and J. Z. are supported by NSF under Grant No. DMR-1506212 and by ONR under Grant No. N00014-11-1-0730. L. Z. T. and S. G. L. are supported by the Theory Program at the Lawrence Berkeley National Lab through the Office of Basic Energy Sciences, U.S. Department of Energy under Contract No. DE-AC02-05CH11231 which provided theoretical analyses and simulations of disorder effects; and by the National Science Foundation under Grant No. DMR15-1508412 which provided for the calculation of electron-electron interaction effects. K. W. and T. T. are supported by the Elemental Strategy Initiative conducted by the MEXT, Japan. T. T. is also supported by a Grant-in-Aid for Scientific Research on Grant 262480621 and on Innovative Areas "Nano Informatics" (Grant No. 25106006) from JSPS. The authors acknowledge use of facilities at the PSU site of NSF NNIN. We are grateful for helpful discussions with X. Hong.

# Supplementary material for

Effective Mass in Bilayer Graphene at Low Carrier Densities: the Role of Potential Disorder and Electron-Electron Interaction

J. Li[1], L. Z. Tan[2,3][†][*], K. Zou[1][‡], A. A. Stabile[1], D. J. Seiwell[1], K. Watanabe[4], T. Taniguchi[4], Steven G. Louie[2,3][*], J. Zhu[1,5][*]

[1]*Department of Physics, The Pennsylvania State University, University Park, Pennsylvania 16802, USA*

[2]*Department of Physics, University of California at Berkeley, Berkeley, California 94720, USA*

[3]*Materials Sciences Division, Lawrence Berkeley National Laboratory, Berkeley, California 94720, USA*

[4]*National Institute for Material Science, 1-1 Namiki, Tsukuba 305-0044, Japan*

[5]*Center for 2-Dimensional and Layered Materials, The Pennsylvania State University, University Park, Pennsylvania 16802, USA*

[†]Current address: *The Makineni Theoretical Laboratories, Department of Chemistry, University of Pennsylvania, Philadelphia, Pennsylvania 19104-6323, USA*

[‡]Current address: *Department of Applied Physics and Center for Research on Interface Structures and Phenomena (CRISP), Yale University, New Haven, Connecticut 06520, USA*

[*]Corresponding to: jzhu@phys.psu.edu (J. Zhu), sglouie@berkeley.edu (S. G. Louie), and liangtan@sas.upenn.edu (L. Z. Tan)

## Outline Supplementary Material

1. SdH oscillation background subtraction
2. Global fitting of $m^*$ and $\tau_q$
3. Estimating disorder density fluctuation from $\sigma(n)$
4. Effect of band gap opening on $m^*$
5. Effect of trigonal warping on $m^*$
6. Fits using the Thomas-Fermi screening
7. Fits including disorder and renormalized TB hopping parameters

# 1. SdH oscillation background subtraction

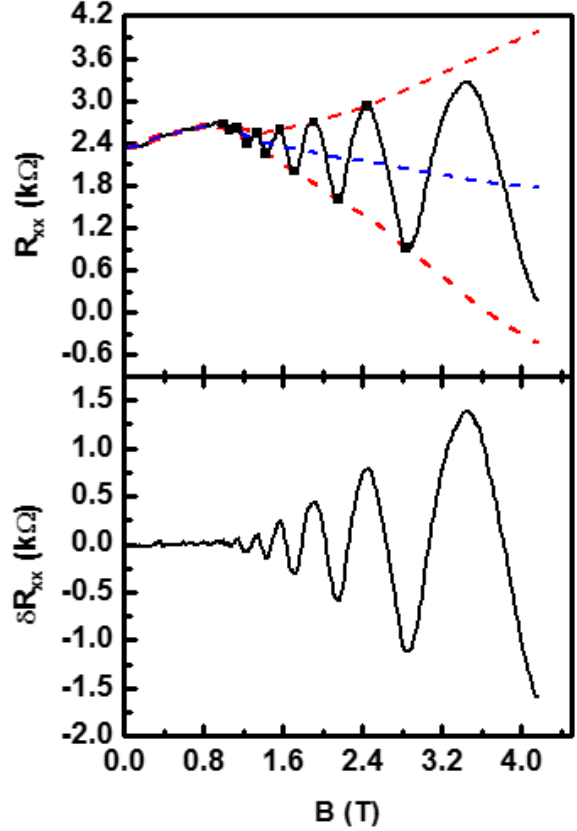

**Figure S1:** Upper: $R_{xx}(B)$ for hole density $n_h = 8.3 \times 10^{11}$ cm$^{-2}$ at $T = 2.3$ K together with the envelopes (red dashed curves) and the calculated background (blue dashed curve). Lower: $\delta R_{xx}(B)$ after the background trace is subtracted. Data from sample B.

Fig. S1 illustrates the background subtraction process. We obtain the upper and lower envelopes of the SdH oscillations (red dashed curves) by using a spline fit to the oscillation maxima and minima (black squares) respectively. The average of the two (blue dashed curve) is subtracted from the $R_{xx}(B)$ data to obtain $\delta R_{xx}$ shown in the lower plot.

# 2. Global fitting procedure

Figure S2 illustrates how we simultaneously determine $m^*$ and $\tau_q$ using the SdH oscillations at multiple temperatures. Fig. S2(a) shows a few exemplary fits at $T = 2.3$ K and 15 K for $n_h = 3.0 \times 10^{11}$ cm$^{-2}$ in sample B. At $T = 2.3$ K, three combinations: $(m^*, \tau_q) = (0.0260\ m_e, 98\ \text{fs})$ or $(0.0285\ m_e, 107\ \text{fs})$, or $(0.0320\ m_e, 120\ \text{fs})$ can all fit data equally well, and the three fits overlap. However, only the pair $m^* = 0.0285\ m_e$ and $\tau_q = 107$ fs can also fit data at $T = 15$ K (Fig. S2(a)) and at all other temperatures (Fig. S2(b)). As trial values of $m^*$ deviate from the optimal value $m^*_0$, systematic deviation of the fit from data guides us towards $m^*_0$ quickly. The cases for $m^* > m^*_0$ and $m^* < m^*_0$ are illustrated in Fig. S2(a).

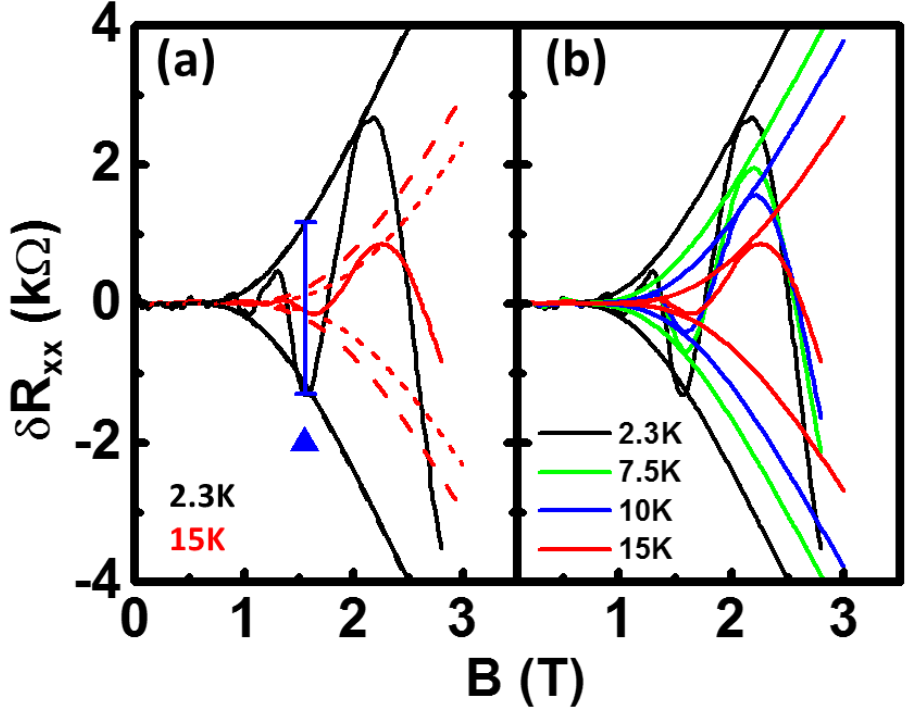

**Figure S2:** Global fitting of $m^*$ and $\tau_q$ to SdH oscillations at all temperatures. (a) Three fitting curves with $m^* = 0.026$ $m_e$ and $\tau_q = 98$ fs (dashed curves), $m^* = 0.0285$ $m_e$ and $\tau_q = 107$ fs (solid curves) and $m^* = 0.032$ $m_e$ and $\tau_q = 120$ fs (short dashed curves). All three sets fit the $T = 2.3$ K data well. Only $m^*_0 = 0.0285$ $m_e$ and $\tau_q = 107$ fs (solid curves) also fit the $T = 15$ K data. (b) Fits using $m^*_0 = 0.0285$ $m_e$ and $\tau_q = 107$ fs describe data at a range of temperatures very well. From sample B. Hole density $n_h = 3.0 \times 10^{11}$ cm$^{-2}$.

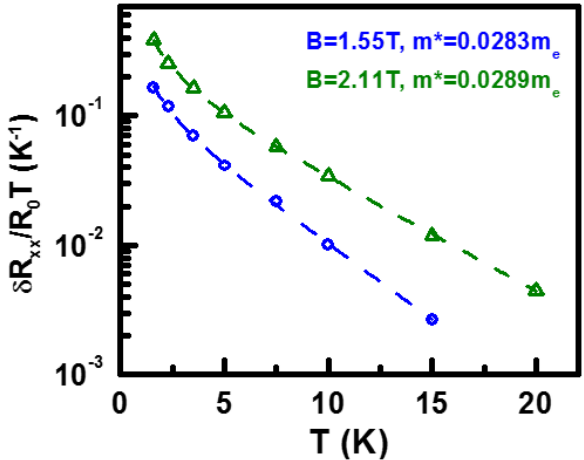

**Figure S3:** $\delta R_{xx}/R_0 T$ versus $T$ in a semilog plot for hole density $n_h = 3.0 \times 10^{11}$ cm$^{-2}$ in sample B at $B = 1.55$ T (circles), and at $B = 2.11$ T (triangles). The dashed curves are fits to Eq. (2) using $m^* = 0.0283$ $m_e$ for $B = 1.55$ T and $m^* = 0.0289$ $m_e$ for $B = 2.11$ T respectively. $\tau_q = 107$ fs for both.

Fig. S3 illustrates how we estimate the uncertainty of $m^*$. Curves generated using the optimal values of $m^*$ and $\tau_q$ (determined by the global fitting) are plotted together with data, as shown in Fig. S2 (b). We read off the temperature-dependent oscillation amplitude $\delta R_{xx}(T)$ at a fixed magnetic field combing measurement at one end and the global fit on the other end (see e.g. B= 1.55 T marked by a triangle in Fig. S2(a)). We then plot $\delta R_{xx}/R_0 T$, where $R_0$ is the zero-field $R_{xx}$, as a function of temperature and fit to Eq. (2) of the main text to obtain $m^*$. Two examples are shown in Fig. S3. We typically do 4 to 6 fittings for each carrier density, and obtain the average and standard deviation of $m^*$. The standard deviation of $m^*$ varies from ± 0.0002 $m_e$ to ± 0.004 $m_e$ from high to low densities.

## 3. Estimating disorder density fluctuation from $\sigma(n)$

Figure S4 plots the sheet conductance versus carrier density in both samples A and B in a log-log plot. Following Ref. [1], we estimate the disorder inducted density fluctuation $\delta n$ by locating the onset carrier density $n^*$, where $\sigma(n)$ starts to increase sharply with $n$. $n^*$ is found to be ~ $2 \times 10^{10}$ cm$^{-2}$ and ~ $4 \times 10^{10}$ cm$^{-2}$ in sample A and B respectively.

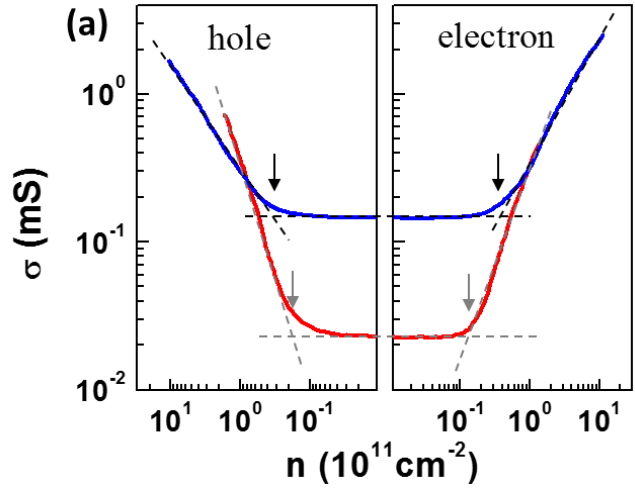

**Figure S4:** (a) Sheet conductance vs carrier density $\sigma(n)$ for samples A (solid red) and B (solid blue) in a log-log plot. The arrows point to $n^*$'s. Dashed lines illustrate how they are determined.

## 4. Effect of band gap opening on $m^*$

In this section, we discuss the effect of an electric field $D$ induced band gap on the bare band mass $m^*_b$. The opening of a band gap leads to the enhancement of $m_b$ at low carrier densities [2], thus can potentially impact the interpretation of the measured $m^*$. In sample B, the residual chemical doping from above and below the bilayer graphene is approximately known, using knowledge from a dual-gated sample in immediate proximity. Following Ref. [3], we can compute the band gap parameter $\Delta$ as a function of the carrier density induced by the backgate accurately, then use this information to compute the tight-binding (TB) bands using a 4 × 4 Hamiltonian and a full set of hopping parameters determined by Zou et al [4]. We then calculate the band mass $m_b$ using Eq. (1) of the main text. Figure S5 (a) and (b) plot the calculated $\Delta$ and $m_b$ as a function of the backgate-induced carrier density. Also plotted for comparison in Fig. S5 (b) is $m_b$ calculated using $\Delta = 0$ (solid lines).

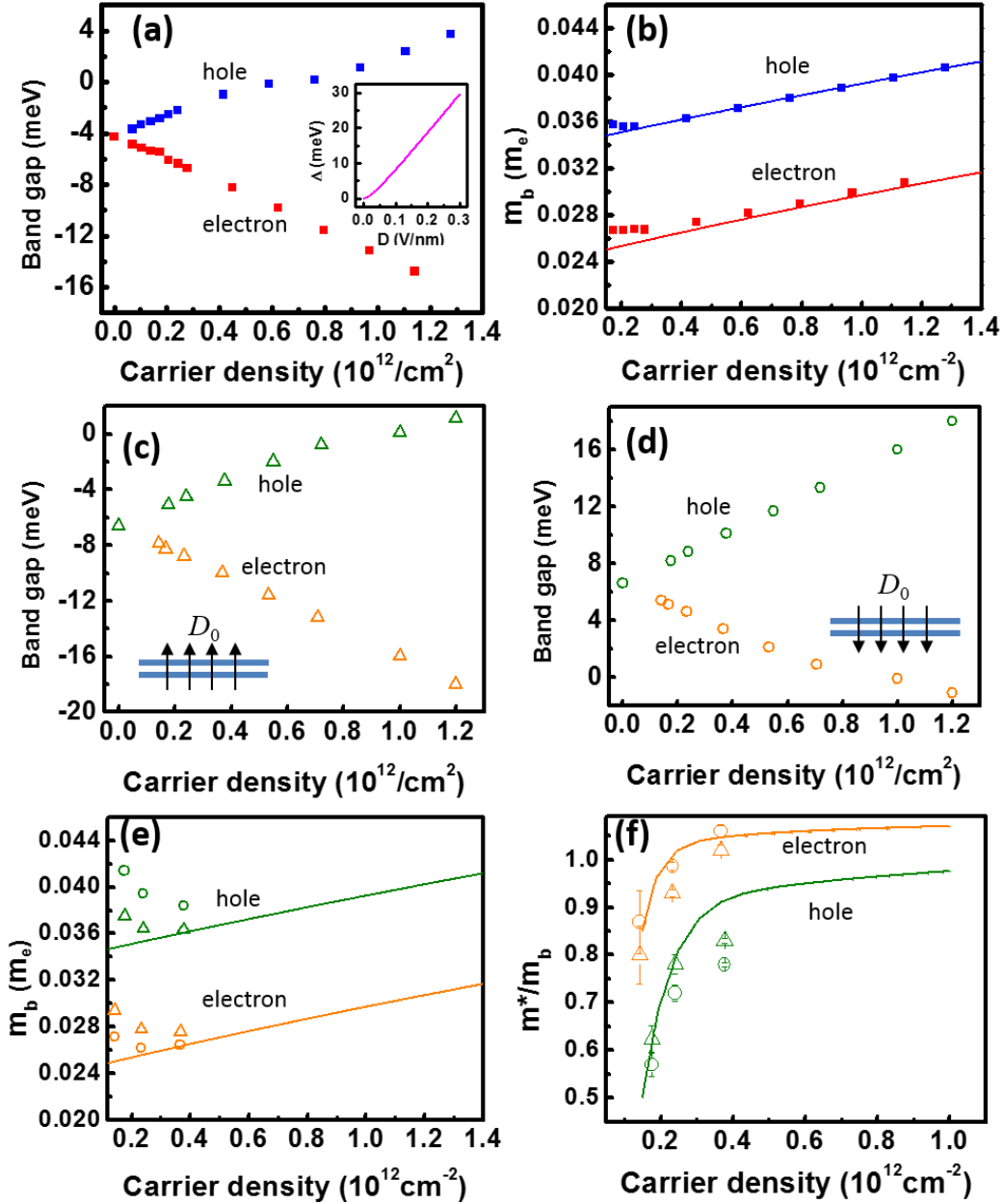

**Figure S5:** (a) and (b) are for sample B. (c)-(f) are for sample A. (a) Calculated band gap $\Delta$ vs hole (blue) and electron (red) density. Inset: calculated $\Delta$ vs displacement field $D$ following formulae given in Ref. [3]. (b) The TB band mass $m_b$ vs carrier density including $\Delta$ (symbols) and setting $\Delta = 0$ (solid lines). The symbols follow (a). (c) and (d) Calculated $\Delta$ vs hole (olive) and electron (orange) density for positive (c) and negative (d) $D_0$ scenarios. (e) The calculated $m_b$ vs carrier density including $\Delta$ (symbols) and setting $\Delta = 0$ (solid lines). (f) The ratio of measured $m^*$ vs the band mass $m_b$ including $\Delta$ (symbols). The solid lines plot the ratio of the fitted $m^*$ in Fig. 4 of the main text vs $m_b$ setting $\Delta = 0$. (c) – (f) use the same color and symbol schemes. The TB parameters used in the calculations are $\gamma_0 = 3.43$ eV, $\gamma_1 = 0.40$ eV, $\gamma_3 = 0$ and $v_4 = 0.063$ determined in Ref. [4]. Disorder broadening $\Gamma$ is set to 3.3 meV, which corresponds to $\tau_q = 100$ fs.

It is clear from the comparison that in sample B the effect of the finite $D$-field on $m_b$ is weak for both electrons and holes. At the lowest studied carrier density $n = 2 \times 10^{11}$ cm$^{-2}$, our calculations show a slight *increase* of $m_b$ of less than 2% for holes and 6% for electrons, whereas experiment showed a *suppression* of 30% in $m^*_h$. Neither the trend nor the relative magnitude of the suppression will change significantly when the effect of $\Delta$ is included. The situation in sample A is more complicated. Since the device is not dual-gated, we cannot determine the $D$ field at the CNP accurately. We estimate, using the resistance peak at the CNP of the sample (~ 300 kΩ) and knowledge from other dual-gated bilayer graphene devices [5], that an unintentional electric field of approximately $D_0 = 85$ mV/nm, which corresponds to a band gap of $\Delta_0 \sim 6.6$ meV (see inset of Fig. S5 (a)), is likely present at the CNP in this sample. We discuss two possibilities separately in Figs. S 5 (c) and (d), where $D_0$ points along or opposite to the backgate induced electric field as the schematics in the figures show. Figures S5 (c) and (d) plot the evolution of the band gap $\Delta$ as a function of the carrier density in each scenario respectively. Figure S5 (e) plots the calculated $m_b$ in both scenarios, together with $m_b$ corresponding to $\Delta = 0$ for reference. The symbols correspond to those of (c) and (d). As expected from the direction of $D_0$, the scenario in Fig. S 5 (c) (triangles) leads to appreciable enhancement of the electron band mass $m^e_b$, while leaving the hole band mass $m^h_b$ nearly intact whereas the scenario in Fig. S5 (d) (circles) produces the opposite effect. All enhancement becomes negligible above $n = 4 \times 10^{11}$ cm$^{-2}$. In Fig. 5(f), we plot the "suppression" factor, i.e., the ratio of the measured $m^*$ versus the $\Delta$-enhanced $m_b$ for both scenarios. Because of the enhancement of $m_b$, our measured $m^*$ now appears to be slightly more "suppressed". For example, the suppression of $m^*_h$ at $n = 2 \times 10^{11}$ cm$^{-2}$ increases from 30% ($\Delta = 0$) to 38 or 43 % when $\Delta$ is included. However, a comparison to the theoretical suppression factor $m^* / m_b (\Delta = 0)$ using $m^*$ shown in Fig. 4 of the main text shows that the theoretical fit using disorder parameter $\delta E$=5.4 meV still provides a good fit to the experimental data, regardless of which scenario of $D_0$ occurs in the sample. The agreement with the electron branch, is in fact better than what's shown in Fig. 4 of the text. This discussion shows that the conclusions of the paper are robust even when the effect of the band gap on $m^*$ is taken into account.

## 5. Effect of trigonal warping on $m^*$

The inter-layer hopping integral $\gamma_3$ deforms the spherical symmetry of the Fermi surface as shown in Fig. S6 (a). This effect becomes more pronounced at low carrier densities and can lead to the breaking up of the Fermi surface into three pockets, i.e. the Lifshitz transition [6]. Although the density range studied here (0.2 ~ 1.2 × 10$^{12}$ cm$^{-2}$) is far above the Lifshitz transition density, we investigated the role of the warping on $m^*$. Figure S6 (a) plots examples of deformed Fermi surfaces for $E_F = 7$ meV and 30 meV respectively for electrons (blue) and holes (red), using a 4 × 4 Hamiltonian (see Fig. S5) and the largest $v_3 = \gamma_3/\gamma_0 = 0.11$, i.e. $\gamma_3 = 0.38$ eV, found in the literature [7-9] $E_F = 7$ meV corresponds to the lowest carrier densities we measured. The corresponding $m^*$ is plotted in Fig. S6 (b), together with calculations corresponding to $\gamma_3 = 0$. It is clear from the comparison that trigonal warping plays a negligible role on $m^*$ in this density range, despite the deformation of the Fermi surface. We have therefore set $\gamma_3 = 0$ in all subsequent calculations.

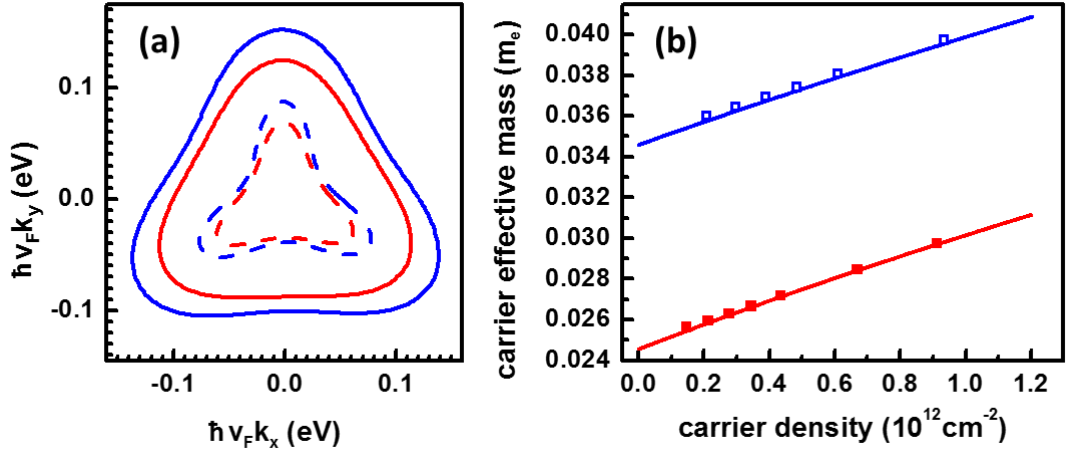

**Figure S6:** (a) Warped Fermi surfaces in momentum space for $E_F = 30$ meV (solid curve) and $E_F = 7$ meV (dashed curve) for holes (blue) and electrons (red). They correspond to $n_h = 2 \times 10^{11}$ cm$^{-2}$, $n_e = 1.5 \times 10^{11}$ cm$^{-2}$ ($E_F = 7$ meV) and $n_h = 9.4 \times 10^{11}$ cm$^{-2}$, $n_e = 6.7 \times 10^{11}$ cm$^{-2}$ ($E_F = 30$ meV) respectively. $v_3 = 0.1$. (b) Calculated electron and hole mass $m^*$ using $v_3 = 0.11$ (symbols) and $v_3 = 0$ (solid curves). Other TB parameters are $\gamma_0 = 3.43$ eV, $\gamma_1 = 0.4$ eV, and $v_4 = 0.063$.

## 6. Fits using the Thomas-Fermi screening

Figure S7 plots the theoretical fit of the data using Thomas-Fermi screening. The Thomas-Fermi wavevector $q_{TF}$ is reduced ten-fold from its expected value to fit the hole mass data. This ten-fold reduction implies extremely weak screening, which cannot be justified in our devices. Even so, the agreement with the electron branch is still poor. Together with the RPA results plotted in Fig. 4 of the main text, these calculations show that e-e interaction alone cannot account for the mass suppression observed in experiment.

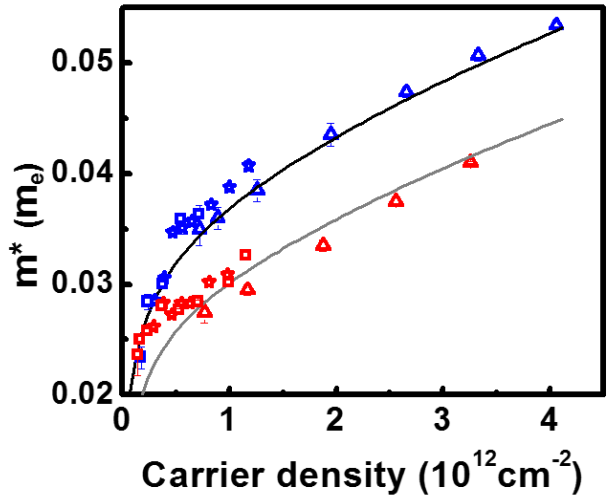

**Figure S7:** $m^*$ calculated using the Thomas-Fermi screened self-energy and the T-F wavevector $q_{TF} = 0.1 m^* e^2 / \hbar^2$. The legend follows Fig. 4 of the main text. Hopping parameters are: $\gamma_0 = 3.08$ eV, $\gamma_1 = 0.37$ eV, $\gamma_3 = 0$, and $v_4 = 0.063$. Dielectric constant of h-BN $\varepsilon_{BN} = 3$.

## 7. Fits including disorder and renormalized TB parameters

In the text, we showed that e-e interaction explicitly accounted for using the RPA approximation + disorder broadening provides an excellent description of data. Here we investigate whether the interaction effect can continue to be represented by renormalized TB parameters, as was done in Zou et al [4] for high carrier densities. Figure S8 plots the calculated $m^*$, using the same set of renormalized TB parameters empirically determined in Ref. [4] and including disorder broadening $\delta E = 5.4$ meV, obtained in the best fit in Fig. 4 of the text. The agreement between theory and experiment is also very good. It should be emphasized however, that this agreement does not mean the effect of the e-e interaction is unimportant but rather it can be well captured by renormalized TB parameters in the entire density range ($10^{11}$ - $10^{12}$/cm$^2$) we studied.

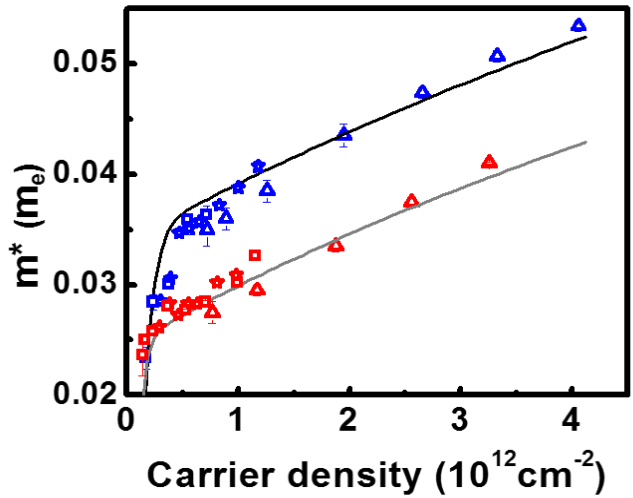

**Figure S8:** $m^*$ calculated using the TB parameters of Ref. [4] and including disorder broadening $\delta E = 5.4$ meV.